\documentclass[aps,prb,showpacs,groupedaddress,twocolumn]{revtex4}

\usepackage{graphicx}
\begin{document}

\title{Persistent spin current in a spin-orbit coupling/normal hybrid ring}

\author{Qing-feng Sun$^{1,2}$, X. C. Xie$^{2,3}$, and Jian
Wang$^{1,\star}$}
\affiliation{ $^1$Department of Physics and the center of theoretical and
computational physics, The University of Hong
Kong, Pokfulam Road, Hong Kong, China;\\
$^2$Beijing National Lab for Condensed Matter Physics and
Institute of Physics, Chinese Academy of Sciences, Beijing 100080,
China;\\
$^3$Department of Physics, Oklahoma State University, Stillwater,
Oklahoma 74078
 }

\date{\today}

\begin{abstract}
We investigate the equilibrium property of a mesoscopic ring with
spin orbit (SO) interaction. It is well known that for a normal
mesoscopic ring threaded by a magnetic flux, the electron acquires
a Berry phase that induces the persistent (charge) current. Similarly,
the spin of electron acquires a spin Berry phase traversing the ring with
SO interaction. It is this spin Berry phase that induces
a persistent spin current. To demonstrate its existence, we calculate
the persistent spin current without accompanying charge current in
the normal region in a hybrid mesoscopic ring. We point out that
this persistent spin current describes the real spin motion and can be
observed experimentally.
\end{abstract}


\maketitle

Recently, physics of semiconductors with the spin-orbit (SO)
interaction has attracted great attention, as it plays an
important role for the emerging field of semiconductor
spintronics.\cite{ref1} SO interaction couples the spin degree of
freedom of electrons to their orbital motions, thereby giving rise
to a useful way to manipulate and control the electron spin by an
external electric field or a gate voltage. Interesting effects
resulting from SO interaction have been predicted. For example,
using the effect of spin precessions due to the Rashba SO
interaction Datta and Das proposed a spin-transistor more than ten
years ago.\cite{ref2} Very recently, a very interesting effect,
the intrinsic spin Hall effect, is theoretically predicted by
Murakami {\it et.al.} and Sinova {\sl et.al.},\cite{ref3,ref4}
that a substantial amount of dissipationless spin current can be
generated from the interplay of the electric field and the SO
coupling. Since then, the spin Hall effect has generated
tremendous interests with a great deal of works focusing in the
field of spintronics.

In this Letter, we explore another interesting effect that a
persistent spin current without accompanying charge current exists
in a coherent mesoscopic semiconductor ring with simplectic
symmetry, i.e., with SO coupling but maintaining the time reversal
symmetry. More than two decades ago, the persistent (charge)
current in a mesoscopic ring threaded by a magnetic flux has been
predicted theoretically,\cite{ref5} and later observed
experimentally in the early 1990s.\cite{ref6} It is now well known
that the persistent charge current is a pure quantum effect and
can sustain without dissipation in the equilibrium case. There has
also been many investigations on the persistent spin
current.\cite{ref7,ref9,ref10,ref11} For example, in a mesoscopic
ring with a crown-shape inhomogeneous magnetic field\cite{ref7} or
threaded by a magnetic flux\cite{ref9}, the persistent spin
current has been predicted and is related to the Berry's phase.
Recently, the persistent spin current carried by Bosonic
excitations has also been predicted in a Heisenberg ring with the
magnetic field or in the ferromagnetic material.\cite{ref10} The
reason that the persistent spin current exists may be explained as
follows. Due to the magnetic field or the magnetic flux, there are
persistent flows of both spin up and down electrons. In the
absence of SO coupling, this gives rise to the well known
persistent charge current. In the presence of SO coupling or
magnetic field, the persistent charge current is spin polarized
resulting a nonzero persistent spin current. Hence the origin of
this persistent spin current is the same as that of persistent
charge current so that the persistent spin current always
accompanies with a persistent charge current.

Up to now, the issue that whether the persistent spin current
without accompanying charge current (a pure persistent spin
current) can be induced solely by SO interaction at zero magnetic
flux or magnetic field has not been addressed.\cite{ref12} In the
present Letter, we show that a non-magnetic semiconducting ring
with SO interaction can sustain a pure persistent spin current in
the absence of the external magnetic field or magnetic flux. Since
the magnetic flux or magnetic field acts like a "driving force"
for the persistent charge current, one naturally looks for the
analogous "driving force" in the spin case. To discuss this
question, let us consider two devices. The first device consists
of a mesoscopic ring (without SO interaction) where a magnetic
atom with a magnetic dipole moment is placed at the center of the
ring (see Fig.1a). In the second device the magnetic atom is
replaced by a charged atom, e.g., an ion (see Fig.1b). The
magnetic atom produces a vector potential ${\bf A}$ on the
perimeter of the ring which drives the persistent charge current.
By analogy, a charged atom which produces a scalar potential
$\phi$ on the perimeter of the same ring should drive a persistent
spin current. Since the presence of this ionic center generates a
SO interaction in the relativistic limit, we expect that SO
interaction which plays the role of the spin "driving force" will
induce a pure persistent spin current.\cite{note2}

The existence of the pure persistent spin current can be examined
from another point of view using Berry phase. It is well known
that an electron circulating a ring with a non-uniform magnetic
field or magnetic flux acquires a geometric phase (Berry
phase).\cite{berry} It has been discovered by Loss et al that it
is this Berry phase $\chi$ that induces the well known persistent
charge current.\cite{ref7} Assuming that the electron wavelength
is much smaller than the perimeter of the ring and the electron
motion is quasi-classical, let us examine an electron with spin
$\sigma$ traverses slowly along the ring with only
SO interaction.\cite{ref19}
Due to the SO interaction, the spin of this electron precesses and
acquires a geometric phase after the electron returns to its
starting point.\cite{ref19} This is the so-called spin Berry
phase\cite{ref19}. The spin Berry phase due to Rashba SO
interaction for an electron with spin $\sigma$ moving in the
clockwise direction is found to be\cite{ref19} $\chi_\sigma =
\sigma\pi$ where $\sigma = \pm$ for $\sigma =
\uparrow,\downarrow$. From the physical picture due to Loss et
al,\cite{ref7} the spin Berry phase $\chi_+$ for the spin up
electron induces a clockwise persistent spin polarized current
$I_1$. Similarly, the Berry phase $\chi_-$ induces a
counter-clockwise persistent spin polarized current with the
polarization exactly opposite to that of $I_1$ since our system
has time reversal symmetry. As a result, the spin Berry phase due
to SO interaction will induce a pure persistent spin current.

Now we present an example to show that indeed a pure persistent
spin current can exist for a semiconducting ring with SO
interaction. In the presence of SO interaction, the spin of an
electron experiences a torque and hence $\sigma_i$ ($i=x,y,z$) is
not a good quantum number anymore. Because of this, the spin
current is not conserved using the conventional definition. At
present there are controversies on whether one should define a
conserved spin current or whether there exists a conserved spin
current.\cite{ref13,ref14} In another word, so far there is no
consensus on the definition for the spin current in the presence
of SO interaction. In the present work, we do not wish to address
the issue of this controversy. We avoid this controversy by
considering a one-dimensional mesoscopic semiconducting ring that
consists of a Rashba SO coupling region and a normal region
without SO interaction as shown in Fig.1c.\cite{note2} Since there
is no spin-flip in this normal region, the spin current can be
defined without controversy.

The Hamiltonian of our system is given by:\cite{ref9,ref15}
\begin{equation}
  H = -E_a \frac{\partial^2}{\partial \varphi^2}
    -\frac{i\sigma_{r}}{2a}\left[
  \alpha_R(\varphi) \frac{\partial}{\partial \varphi}
 + \frac{\partial}{\partial \varphi} \alpha_R(\varphi)
 \right]
  -i \frac{\alpha_R(\varphi)}{2a}\sigma_{\varphi}
\end{equation}
where $E_a =\hbar^2/2ma^2$, $a$ is the radius of the ring, $m$ is
the effective mass of the electron, $\sigma_r = \sigma_x
\cos\varphi +\sigma_y \sin\varphi$, and $\sigma_{\varphi} =
-\sigma_x \sin\varphi +\sigma_y \cos\varphi$. $\alpha_R(\varphi)$
is the strength of the Rashba SO interaction,
$\alpha_R(\varphi)=0$ while $0 <\varphi <\Phi_0$, i.e. in the
normal region, and $\alpha_R(\varphi)$ is a constant $\alpha_R$ in
the SO coupling region with $\Phi_0 <\varphi <2\pi$.

The eigenstates of Hamiltonian (1) can be solved numerically in
the following way.  First in the Rashba SO coupling region
($\alpha_R \not= 0$), the equation $H\Psi(\varphi)=E
\Psi(\varphi)$ has four independent solutions
$\Psi^{SO}_{i}(\varphi)$ ($i=1,2,3,4$):\cite{ref9}
\begin{equation}
 \Psi^{SO}_{1/2}(\varphi) = \left(\begin{array}{l}
       \cos (\theta/2) e^{ik_{1/2}\varphi} \\
       -\sin (\theta/2) e^{i(k_{1/2}+1)\varphi}
       \end{array}
       \right),
\end{equation}
and $\Psi^{SO}_{3/4} = \hat{T} \Psi^{SO}_{1/2}$ with $\hat{T}$
being the time-reversal operator. In Eq.(2), the wave vectors
$k_{1/2}= -1/2 +1/(2\cos\theta) \pm (1/2) \sqrt{
(1/\cos^2\theta)-1 +4E/E_a}$, and the angle $\theta$ is given by
$\tan(\theta)= \alpha_R/(a E_a)$. Similarly, in the normal region
($0 <\varphi <\Phi_0$), the Sch$\ddot{o}$dinger equation has four
independent solutions: $\Psi^N_{1}(\varphi) = (1,0)^{\dagger}
e^{ik\varphi}$, $\Psi^N_{2}(\varphi) = (1,0)^{\dagger}
e^{-ik\varphi}$, and $\Psi^{N}_{3/4} = \hat{T} \Psi^{N}_{1/2}$.
Secondly, the eigen wave function $\Psi(\varphi)$ with the eigen
energy $E$ can be represented as:
\begin{eqnarray}
  \Psi(\varphi) = \left\{ \begin{array}{l}
            \sum\limits_{i} a_i \Psi^{N}_i(\varphi) ,   \hspace{6mm}
             while \hspace{2mm} 0 < \varphi < \Phi_0 \\
            \sum\limits_{i} b_i \Psi^{SO}_i(\varphi) ,   \hspace{5mm}
             while \hspace{2mm} \Phi_0 < \varphi < 2\pi
             \end{array} \right.
\end{eqnarray}
where $a_i$ and $b_i$ ($i=1,2,3,4$) are constants to be determined
by the boundary conditions at the interfaces $\varphi=0$ and
$\Phi_0$.  Here the boundary conditions are the continuity of the
wave function $\Psi(\varphi)|_{\varphi=0^+/\Phi_0^+} =
\Psi(\varphi)|_{\varphi=2\pi^-/\Phi_0^-} $ and the continuity of its
flux $\hat{v}_{\varphi} \Psi|_{\varphi=0^+/\Phi_0^+} =
 \hat{v}_{\varphi} \Psi|_{\varphi=2\pi^-/\Phi_0^-} $, where
$\hat{v}_{\varphi} =a (\partial \varphi/\partial t) \sim
\partial/\partial \varphi +(i/2) \sigma_r
\tan(\theta) $ is the velocity operator. By using the boundary
conditions, we obtain eight series of linear equations with the
variables $\{a_i, b_i\}$. Then, by setting the determinant of the
coefficients to be zero, the eigenvalues $E_n$ are obtained
numerically. Fig.1d shows $E_n$ versus the Rashba SO strength
$\alpha_R$. For the normal ring ($\alpha_R =0$), the eigenvalues
are $n^2 E_a$ with fourfold degeneracy, and the corresponding
eigenstates are $(1,0)^{\dagger} e^{\pm i n \varphi}$ and
$(0,1)^{\dagger} e^{\pm i n \varphi}$. As the SO interaction is
turned on the degenerate energy levels split while maintaining
twofold Kramers degeneracy. The higher the energy level, the
larger this energy split. Typically, the splits are on the order
of $E_a$ at $\alpha_R=10^{-11}eVm$, with $E_a \approx 0.42meV$ for
the ring's radius $a=50nm$ and the effective mass $m=0.036m_e$.
The eigenvalues $E_n$ versus the normal region's angle $\Phi_0$
also are shown (see Fig.1f). For $\Phi_0 =2\pi$, the whole ring is
normal and $E_n$ are fourfold degenerate. When $\Phi_0$ is away
from $2\pi$, the degeneracy are split into two, and the splits are
larger with the smaller $\Phi_0$. When $\Phi_0 =0$, the whole ring
has the Rashba SO interaction, and the split reaches the maximum.

Since $E_n$ is twofold degenerated, we obtain two eigenstates from
Eqs.(4-10) for each $E_n$, which are labeled $\Psi_n(\varphi)$ and
$\hat{T}\Psi_n(\varphi)$. With the wave functions, the spin
current contributed from the level $n$ can be calculated
straightforwardly by using $I^n_{Si}(\varphi) = Re
\Psi_n^{\dagger} \hat{v}_{\varphi} \hat{\sigma}_i \Psi_n$
($i=x,y,z$). Since there is a controversy about the definition of
spin current in the SO region, we will calculate the spin current
only in the normal region. We note that the spin current is
conserved in the normal region independent of the angle coordinate
$\varphi$. Fig.2 shows the spin current $I^n_{Si}$ versus the
Rashba SO strength $\alpha_R$ for $\Phi_0 =\pi$ (i.e. a half of
the ring is normal and another half has the SO interaction). Our
results show that $I^n_{Sx}$ is exactly zero for all level $n$,
and $I^n_{Sy}$ and $I^n_{Sz}$ exhibit the oscillatory pattern with
$\alpha_R$. A $\pi/2$-phase shift between $I^n_{Sy}$ and
$I^n_{Sz}$ is observed with $\sqrt{(I^n_{Sy})^2 +(I^n_{Sz})^2}$
approximately constant. For two adjacent levels $2n-1$ and $2n$,
their spin current have opposite signs, and $I^{2n-1}_{Si} +
I^{2n}_{Si}=0$ if $\alpha_R=0$. We note that the spin current
$I^n_{Si}$ is quite large. For example, the value $E_a $ is
equivalent to the spin current of a moving electron in the ring
with its speed $4\times 10^5 m/s$.

Now we calculate the equilibrium total spin current $I_{Si}$
contributed from all occupied energy levels: $I_{Si} = 2\sum_n
I^{n}_{Si} f(E_n)$, where $f(E_n) =1/\{\exp[(E_n-E_f)/k_B T]+1\}$
with the Fermi energy $E_f$ and the temperature $T$. The factor
$2$ is due to the Kramers degeneracy. The persistent charge
current and the equilibrium spin accumulation are found to be zero
because the system has the time-reversal symmetry and the states
$\Psi_n$ and $\hat{T} \Psi_n$ have completely opposite charge
current and the spin accumulation. Fig.3a,b show the total spin
currents $I_{Si}$ versus the Rashba SO strength $\alpha_R$ for
different $E_f$. One of the main results is that the spin current
indeed is non-zero when $\alpha_R \not=0$. The persistent spin
currents $I_{Si}$ in Fig.3 have the following features. At
$\alpha_R =0$ the whole ring is normal, so $I_{Si}$ is exactly
zero. With $\alpha_R$ increasing, $I_{Si}$ first increases and it
then oscillates for the large $\alpha_R$. At certain $\alpha_R$,
there is a jump in the curve of $I_{Si}$ versus $\alpha_R$.
Because for this $\alpha_R$ the Fermi energy $E_F$ is in line with
a level $E_n$, leading to a change of its occupation. At zero
temperature, the jump is abrupt as shown in Fig.3a,b. But at
finite temperature, this jump will be slightly smooth. In fact,
these results are similar with the persistent (charge) current in
the mesoscopic ring.\cite{ref5}

The spin current $I_{Si}$ versus the angle of normal region
$\Phi_0$ at a fixed $\alpha_R =3\times 10^{-11} eVm$ is shown in
Fig.3c. While $\Phi_0 =2\pi$, the whole ring is normal, and
$I_{Sx/y/z}=0$. When $\Phi_0$ is away from $2\pi$, the spin
current $I_{Sx/y/z}$ emerge. There perhaps exist the jump in the
curve $I_{Si}$-$\Phi_0$ which behaviors is similar as the jump in
the curve $I_{Si}$-$\alpha_R$. In particular, while $\Phi_0$ tends
to zero, i.e. the normal region tends vanishing, the spin current
$I_{Sx}$ and $I_{Sz}$ still exist. This means that the normal
region is not necessary for generating $I_{Si}$.

From our results several observations are in order: (i) Since the
spin current is calculated in the normal region with no spin-flip,
the spin current is conserved. Hence the existence of spin current
in the present system is not due to the definition of the spin
current. We consider that this spin current describes the real
motion of spins. In addition, the spin current should also exist
in the SO coupling region. (ii) The present system is non-magnetic
so this spin current is solely due to the SO interaction. (iii)
This spin current calculated from the eigenstate of the system is
an equilibrium property of the system. It can sustain without
dissipation in the equilibrium case, i.e. it is a persistent spin
current. (iv) Besides the ring, the device can also be of other
shapes, e.g. a disc device or a quasi one-dimension quantum wire,
and so on. So it is a generic feature that a pure persistent spin
current appears in the SO coupling semiconductor as long as the
size of the device is within the phase coherence length. (v) It is
well known that the persistent charge current can generate a
magnetic field. Similarly, the persistent spin current can
generate an electric field, which offers a way of detecting
it.\cite{ref10,ref18}

In summary, we show that a persistent spin current without
accompanying charge current exists in a semiconducting ring with
only spin-orbit (SO) interaction so that the time reversal
symmetry is retained. It is the spin Berry phase due to the SO
interaction that induces the pure persistent spin current in the
ring. We demonstrate the existence of the persistent spin current
in normal region of a SO coupling/normal hybrid device. We point
out that this spin current describes the real spin motion and can
be measured experimentally. Our persistent spin current in a
semiconducting ring with SO interaction is an analog of the
persistent charge current in the mesoscopic ring threaded by
magnetic flux.

{\bf Acknowledgments:} We gratefully acknowledge the financial
support from a RGC grant from the Government of HKSAR grant number
HKU 7044/05P (J.W.); the Chinese Academy of Sciences and NSF-China
under Grant Nos. 90303016, 10474125, and 10525418 (Q.F.S.); US-DOE
under Grant No. DE-FG02-04ER46124 and NSF under CCF-052473
(X.C.X.).

\newpage

\begin{figure}
\caption{ (Color online) (a) and (b) are the schematic diagrams
for a mesoscopic ring with a magnetic atom or an ion at its
center. (c) Schematic diagram for a hybrid mesoscopic ring having
Rashba SO interaction in part of the ring and the other part being
normal. (d) and (f) show the eigen energies $E_n$ vs. $\alpha_R$
for $\Phi_0=\pi$ and vs. $\Phi_0$ for $\alpha_R=3\times 10^{-11}
eVm$, respectively. The ring's radius $a=50nm$. } \label{fig:1}
\end{figure}

\begin{figure}
\caption{ $I^n_{Sy}$ (a) and $I^n_{Sz}$ (b) vs $\alpha_R$ for
$\Phi_0=\pi$ and $a=50nm$. Along the arrow direction, $n=7$, $5$,
$3$, $1$, $2$, $4$, $6$, and $8$.
 }
\label{fig:2}
\end{figure}

\begin{figure}
\caption{ (Color online) (a) and (b) show $I_{Sy}$ and $I_{Sz}$ vs
$\alpha_R $ for $\Phi_0=\pi$. (c) shows $I_{Sx/y/z}$ vs the normal
region's angle $\Phi_0$ for $\alpha_R=3\times 10^{-11} eVm$ and
$E_f=3E_a$. The solid, dashed, and dotted lines correspond to
$I_{Sx}$, $I_{Sy}$, and $I_{Sz}$, respectively. The ring's radius
$a=50nm$ and the temperature $T=0$.
 } \label{fig:3}
\end{figure}


\begin{references}
\bibitem{note2}
Consider the case that there is a point charge at the center of
the ring that gives rise to a Thomas SO interaction. We find that
the persistent spin current also exists, similar as for the Rashba
SO interaction case discussed in the text.
\end{references}
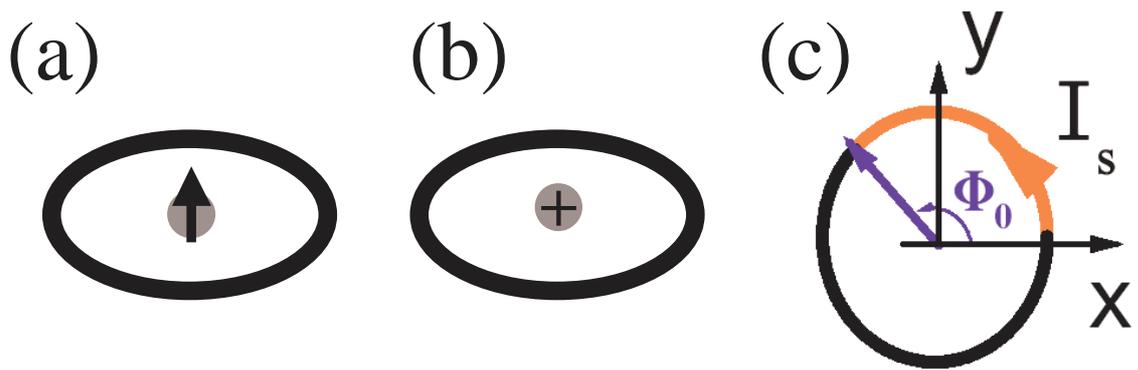
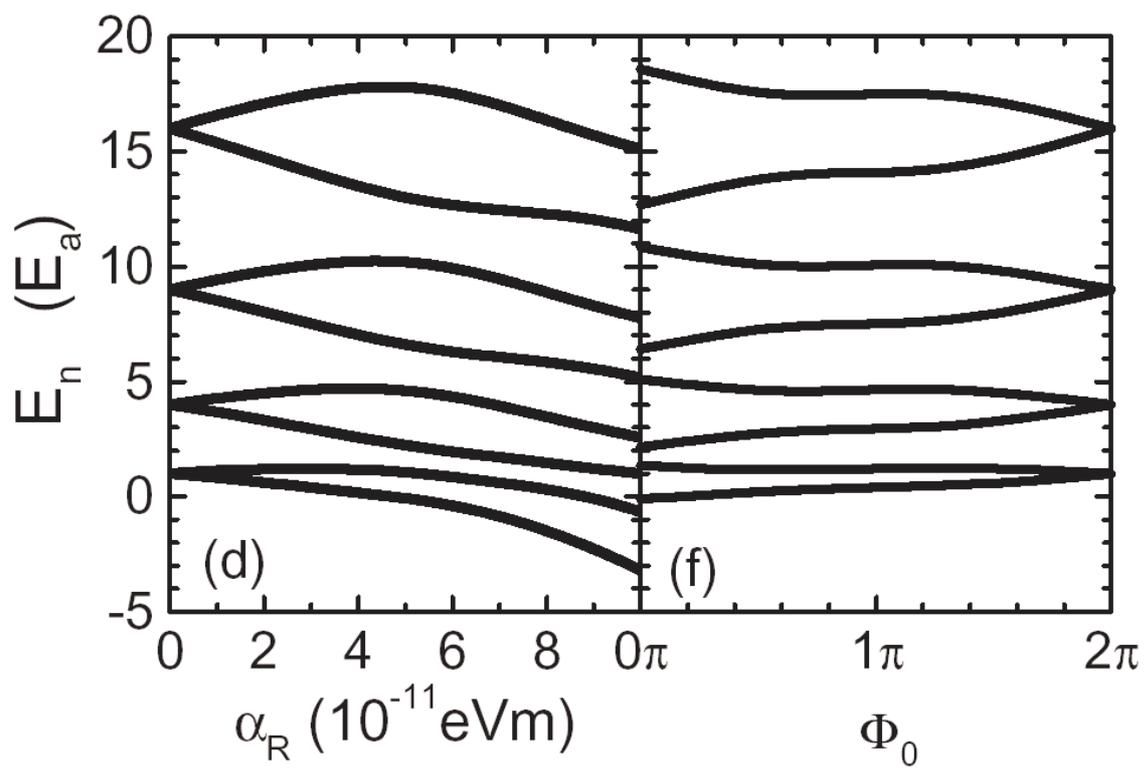
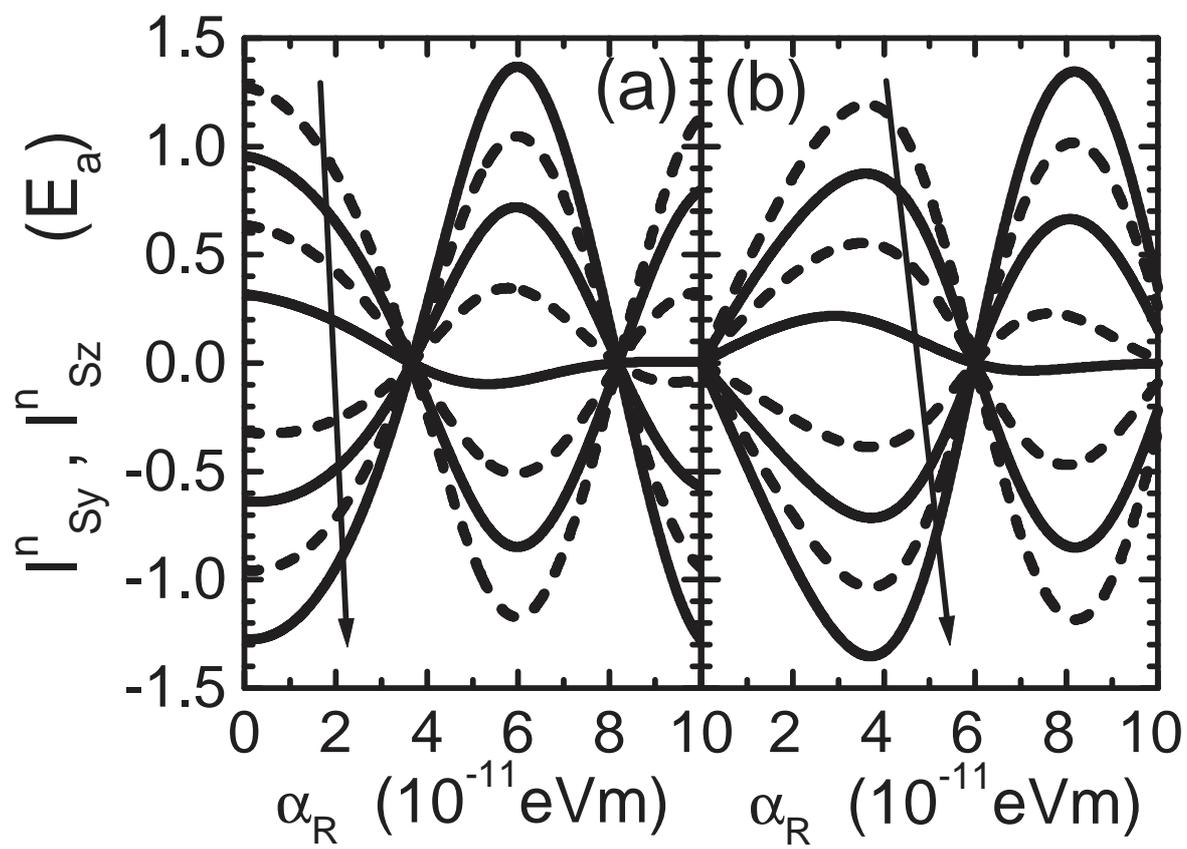
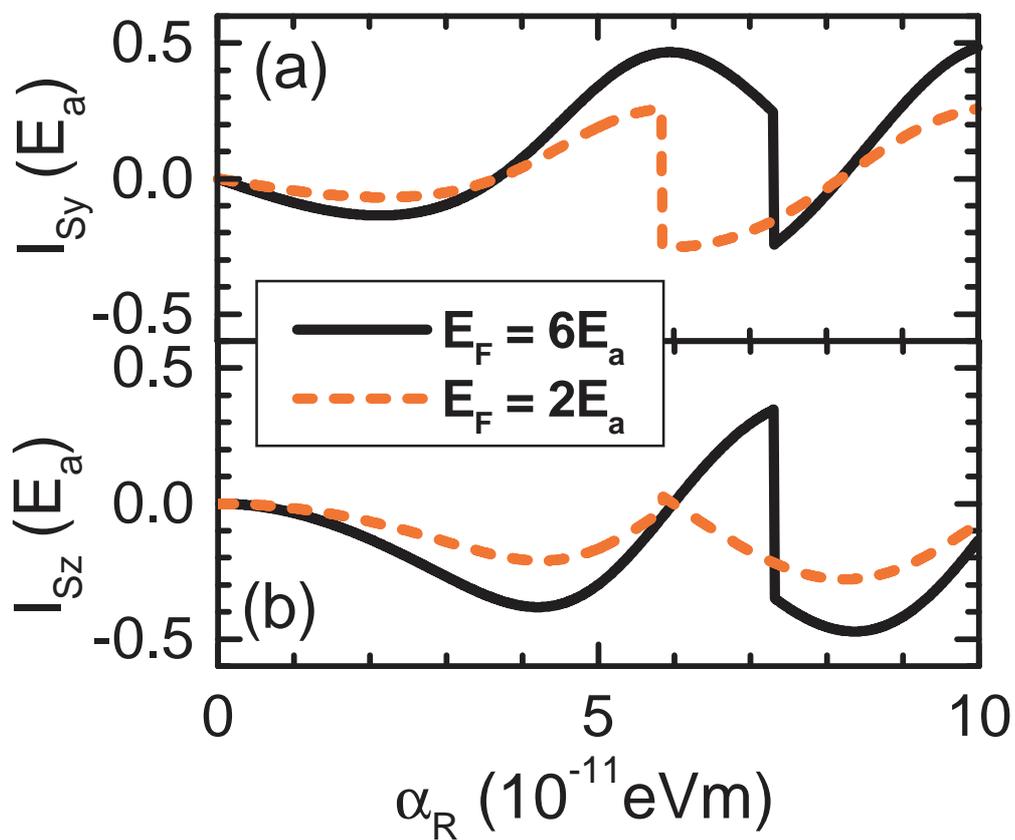
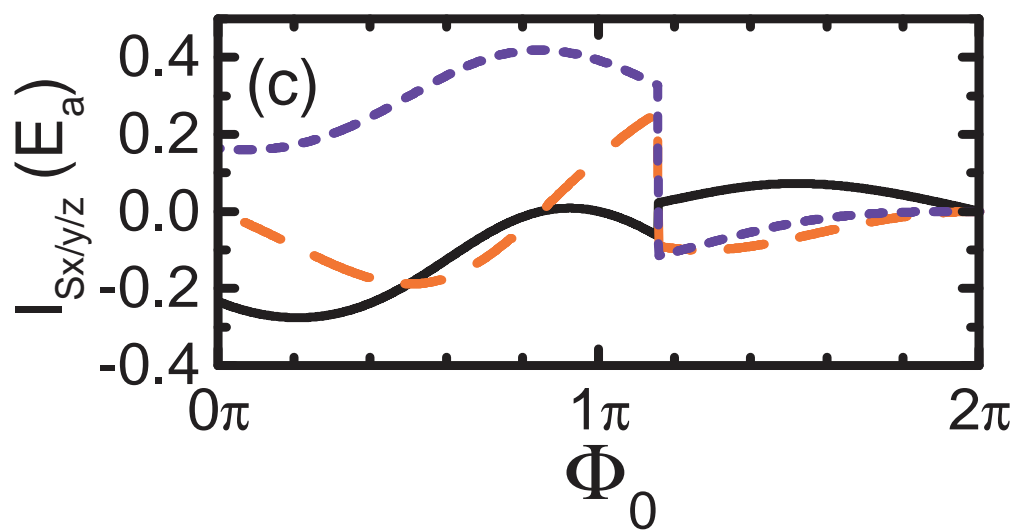
\end{document}